\documentclass[prb,twocolumn,amsmath,amssymb,floatfix,showpacs,superscriptaddress]{revtex4}

\usepackage{graphicx}

\usepackage{amssymb}
\usepackage{color}
\usepackage{epsfig}

\begin{document}

\def\gammav{{\mbox{\boldmath{$\gamma$}}}}
\def\sigmav{{\mbox{\boldmath{$\sigma$}}}}

\title{Spin filtering by a periodic nanospintronic device}

\author{Amnon Aharony}

\altaffiliation{Also at Tel Aviv University.}

\affiliation{Department of Physics and the Ilse Katz Center for
Meso- and Nano-Scale Science and Technology, Ben Gurion
University, Beer Sheva 84105, Israel}

\author{Ora Entin-Wohlman}

\altaffiliation{Also at Tel Aviv University.}

\affiliation{Department of Physics and the Ilse Katz Center for
Meso- and Nano-Scale Science and Technology, Ben Gurion
University, Beer Sheva 84105, ISRAEL}

\author{Yasuhiro Tokura}

\affiliation{ NTT Basic Research Laboratories,
NTT Corporation, Atsugi-shi, Kanagawa 243-0198, Japan}

\author{Shingo Katsumoto}

\affiliation{Institute of Solid State Physics, University of Tokyo, Kashiwa, Chiba 277-8581, Japan}

\date{\today}

\begin{abstract}
 For a linear chain of diamond-like elements,
we show that the Rashba spin-orbit interaction (which can be tuned
by a perpendicular gate voltage) and the Aharonov-Bohm flux (due
to a perpendicular magnetic field) can combine to select only one
propagating ballistic mode, for which the electronic spins are
fully polarized along a direction that can be controlled by the
electric and magnetic fields and by the electron energy. All the
other modes are evanescent. For a wide range of parameters, this
chain can serve as a spin filter.
\end{abstract}
 \maketitle

\section{Introduction}

In addition to their charge, electrons also carry a spin, which is
the quantum relativistic source for the electron's intrinsic
magnetic moment. Future device technology and quantum information
processing may be based on spintronics, \cite{1} where one
manipulates the electron's spin (and not only its charge).   One
major aim of spintronics is to build mesoscopic spin valves (or
spin filters), which generate a tunable spin-polarized current out
of unpolarized sources. Much recent research aims to achieve this
goal by using narrow-gap semiconductor heterostructures, where the
spins are subject to the Rashba \cite{3} spin-orbit interaction
(SOI): in a two-dimensional electron gas confined by an asymmetric
potential well, the strength of this SOI can be varied by an
electric field perpendicular to the plane in which the electrons
move. \cite{koga} An early proposal of a spin field-effect
transistor \cite{2} used the Rashba SOI to control the spin
precession of electrons moving in quasi-one-dimensional wires.
Placed between two ferromagnets, the transport of polarized
electrons through such a semiconductor could be regulated by the
electric field. However, such devices are difficult to make, due
to the metal-semiconductor conductivity mismatch.

\begin{figure}[ht]
\vspace{-.5cm}
\begin{center}
\includegraphics[width=5.6 cm]{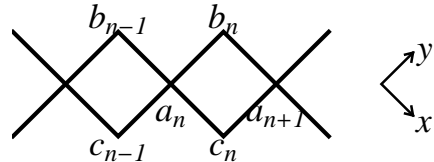}
\end{center}
\vspace{-1.cm} \caption{Chain of diamonds.}\label{1}
\end{figure}

Some of the most striking quantum effects arise due to
interference, which is best demonstrated in quantum networks
containing loops. Indeed, interference due to the Rashba SOI has
been measured on a nanolithographically-defined square loop array.
\cite{koga06} Here we discuss the possibility to construct a spin
filter from such loops. Recently, several groups proposed spin
filters based on a {\it single} loop, subject to both an electric
and a magnetic [Aharonov-Bohm (AB) \cite{AB}] perpendicular
fields. \cite{citro,hatano,oreg} However, such devices produce a
full polarization of the outgoing electrons only for {\it special
values} of the two fields. In the present paper we consider a
chain of such loops, as shown in Fig. \ref{1}. The effects of the
Rashba SOI on the spectrum of the diamond chain of Fig. \ref{1}
was studied by Bercioux {\it et al.}. \cite{berc1} They found a
strong variation of the averaged (over energies) conductance with
the strength of the SOI, which they associated with localization
of the electron due to interference between different paths in
each diamond. Later, this group \cite{berc2} found similar effects
due to both the SOI and an AB flux. However, the possibility to
use such networks to achieve spin filtering has not been
considered. As we show below, the polarization of the outgoing
electrons depends on the energy. Therefore, averaging over
energies mixes different polarization directions and eliminates
the possibility of obtaining full polarization.

 We find that both the ballistic conductance and the spin polarization of the
electrons going through the device can be sharply varied by an
electric field (determining the SOI \cite{koga}), a magnetic field
(determining the AB phases of the orbital electronic wave
functions) and the electrons' energy (set by the chemical
potential in the source). Varying these three parameters, we find
large parameter ranges where all the
energy eigenstates of the device except one become evanescent and
decay exponentially, forming the localized states discussed in
Refs. \onlinecite{berc1} and \onlinecite{berc2}. However, the
electrons in the remaining single mode propagate with {\it fully
polarized spins}. Thus, electrons which enter with arbitrary spins
exit fully polarized. Since this polarization can be tuned by the
parameters, our system is an ideal spin filter.

Section II outlines the tight binding model which we use for solving the Schr\"odinger equation on the periodic chain of diamonds.
Section III presents results for the ballistic conductance and for the polarization of the electrons in the regions where they are fully polarized.
Finally, Sec. IV contains a discussion of our results, including a comparison with the case of a single diamond and a discussion of the application of our results to a finite chain.

\section{Tight binding model}

With SOI, we need to solve for the two-component spinor at each
point on the network.  Bercioux {\it et al.} \cite{berc1,berc2}
treated each bond of the network as a continuous one-dimensional
(1D) wire.
Having expressed the solutions along each bond in terms of the
spinors of the nodes at its two ends, they used the Neumann
 boundary conditions at the nodes to derive
discrete equations for the spinors at these nodes. As we discuss
elsewhere, \cite{deG} these boundary conditions are sufficient but
not necessary for current conservation at the nodes. A more
systematic way to treat such network replaces each continuous bond
bond by a discrete sequence of sites, and then studies the tight
binding model for the wave functions on these sites (and on the
original nodes). As the number of these intermediate sites
increases, one has more sites per unit cell, and therefore one
ends up with more energy bands for the solutions which contain
waves moving along the main axis of the network [i.e. along the
(1,1,0) direction in Fig. \ref{1}]. Qualitatively, we find that
all these bands are similar to each other, and also similar to
those found for the continuous network used in Refs.
\onlinecite{berc1} and \onlinecite{berc2}. Therefore, we choose to
report here only on the simplest case, with no intermediate sites
within the bonds. Thus, we treat a simple tight-binding model,
with sites $\{ u\}$ only on the corners of the diamonds. The
latter model could also describe a network of quantum dots or
anti-dots, located at these nodes. \cite{kats} The stationary
spinors $\Psi_u$, with energy $\epsilon$, obey the Schr\"odinger
equations,
\begin{align}
i \hbar (\partial/\partial t) \Psi_u=\epsilon \Psi_u=-J\sum_v
U^{}_{uv}\Psi_v,\label{TB}
\end{align}
where the sum is over the nearest-neighbor nodes $\{ v\}$, $J$ is
the (real) hopping matrix element (in the absence of fields) and
\begin{align}
U^{}_{uv}\equiv
U^\dagger_{vu}=\exp[i(\phi^{AB}_{uv}+\phi^{SO}_{uv})]
\end{align}
 is a
unitary $2\times 2$ matrix, representing the phase factors due to
the AB flux and to the SOI, $\phi^{AB}_{uv}$ and $\phi^{SO}_{uv}$
respectively. For our structure, all 
bonds are in the $xy-$plane, and  both the uniform magnetic field
${\bf H}=H{\bf {\hat z}}$ and the potential asymmetry which
creates the SOI are along the ${\bf z}$-axis. As can be seen from
Fig. \ref{1}, the $n$'th unit cell contains three sites, and Eq.
(\ref{TB}) reduces to equations for the related spinors,
$\Psi_a(n),~\Psi_b(n)$ and $\Psi_c(n)$. Choosing the edges of the
diamonds along the $x-$ and $y-$ axes (see Fig. \ref{1}), so that
site $a_n$ is located at ${\bf r}_n=(n,n,0)L$ ($L$ is the length
of each edge), the unitary hopping matrices within the $n$'th
diamond are given by \cite{TBSOI}
\begin{align}
&U^{}_{ab}(n)=U^\dagger_{ba}(n)\equiv e^{in\phi/2}e^{i\alpha\sigma_x},\nonumber\\
&U^{}_{ac}(n)=U^\dagger_{ca}(n)\equiv e^{-in\phi/2}e^{-i\alpha\sigma_y},
\end{align}
where ${\bf \sigmav}$ is the vector of Pauli matrices,
$\alpha=k^{}_{SO}L$ ($k^{}_{SO}$ measures the strength of the
`microscopic' SOI, $(\hbar/m)k^{}_{SO}\sigmav \times{\bf p}$) and
$\phi=2\pi H L^2/\Phi_0$ represents the AB phase associated with a
single square diamond (here, $\Phi_0=hc/e$ is the flux unit; $h$
is Planck's constant, $c$ is the speed of light and $e$ is the
electron charge). Note that the dependence of $U^{}_{ab}(n)$ and of $U^{}_{ac}(n)$ on $n$
results from our choice of gauge for the vector potential. The net flux through each diamond is equal to $\phi$, independent of $n$.

For $\epsilon=0$ one encounters dispersionless modes, for which
$\Psi_a(n)\equiv 0$. Since these solutions have zero velocity, and
therefore carry no current, we ignore them in the following
discussion. We next eliminate the spinors $\Psi_b(n)$ and
$\Psi_c(n)$ from the equations, and end up with effective
one-dimensional equations,
\begin{align}
4\Lambda\Psi_a(n)={\bf V}^\dagger\Psi_a(n-1)+{\bf
V}\Psi_a(n+1),\label{ren}
\end{align}
with $4\Lambda=(\epsilon/J)^2-4$ and
\begin{align}
&{\bf V}=U_{ab}(n)U_{ac}(n+1)+U_{ac}(n)U_{ab}(n+1)\nonumber\\
&=e^{-i\phi/2}e^{i\alpha\sigma_x}e^{-i\alpha\sigma_y}
+e^{i\phi/2}e^{-i\alpha\sigma_y}e^{i\alpha\sigma_x}.
\label{V}
\end{align}
Unlike the individual $U_{uv}$'s, the `renormalized'
hopping matrix ${\bf V}$ is {\it not} unitary. This lack of
unitarity reflects interference between the two paths in a
diamond, which may decrease the current along the chain.

In the following we concentrate on propagating waves, 
\begin{align}
&\Psi_a(n)=Ae^{iq{\bar L}n}\chi_a(q), \label{PSI} \end{align}
 where
${\bar L}=L\sqrt{2}$ is the lattice constant of the diamond system
along its axis $(1,1,0)$, the (real) wave-vector $q$ is in the
range $-\pi/2<q{\bar L}<\pi/2$ and $\chi_a$ is a normalized spinor
(which depends on $q$). For such solutions, Eq. (\ref{ren})
implies that $\chi_a$ must obey the  eigenvalue equation  ${\cal
H}\chi_a(q)=\Lambda\chi_a(q)$, with the $2\times 2$ hermitian
matrix
\begin{align}
4{\cal H}=e^{-iq{\bar L}}{\bf V}^\dagger+e^{iq{\bar
L}}{\bf V}.
\end{align}
 We next write
 \begin{align}
 {\cal H}=A+{\bf B}\cdot \sigmav,
 \end{align}
with
\begin{align}
&A=c^2 \cos(q{\bar L})\cos(\phi/2),\nonumber\\
&{\bf B}=-c s \sin(q{\bar L})\cos(\phi/2)\nonumber\\
&\times \bigl
(1,-1,-\cot(q{\bar L})\tan(\phi/2)s/c \bigr ),
\end{align}
 where
$c=\cos\alpha,~s=\sin\alpha$. It follows that the spinor
$\chi_a(q)$ must be an eigenvector of the spin component along
${\bf n}\equiv {\bf B}/|{\bf B}|$:
${\bf n}\cdot\sigmav \chi_a(q)=\mu \chi_a(q),\ \mu=\pm 1$.
Thus, $\Lambda=A+\mu |{\bf B}|$. Given $\Lambda$, this equation
can be written as a quadratic equation in $x=\cos(q{\bar L})$.
Denoting the solutions by $x^{}_{1,2}$, we end up with four
solutions $q^{\pm}_{1,2}=\pm \arccos x^{}_{1,2}$. These solutions
are propagating (evanescent) if $q$ is real (complex). For each
$q$ one then has $\mu=(\Lambda-A)/|{\bf B}|$, so that $\mu$ is invariant under
flipping the sign of $q$.

Since $\chi_a(q)$ is an eigenvector of ${\bf n}\cdot\sigmav$, each
solution with a given $q$ is associated with a full polarization along the direction ${\bf n}$,
\begin{align}
{\bf S}\equiv \langle \chi_a(q)|\sigmav|\chi_a(q)\rangle = \mu
{\bf n}.\label{spin}
\end{align}
 As usual for Rashba SOI, ${\bf n}$ is always perpendicular to the
direction of motion along the axis of the diamond chain,
$(1,1,0)$. In the absence of an AB flux (i.e. $\phi=0$) ${\bf n}$
remains in the direction $(1,-1,0)$. However, the orbital AB flux
causes a rotation of the polarization axis towards the
${\bf z}-$direction. 
Below we present results for $S_z$ and for $S_{xy}\equiv
(S_x-S_y)/\sqrt{2}$. Since $n^{}_{x,y}$ ($n^{}_z)$ is odd (even)
in $q$, flipping the sign of $q$ flips the sign of $S_{xy}$ but
not that of $S_z$.

The probability current from site $u$ to site $v$ is
\begin{align}
I(u\rightarrow v)=(2J/\hbar)\Im
\langle\Psi_u|U_{uv}|\Psi_v\rangle.
\end{align}
The current from site $a_n$
to site $a_{n+1}$ on the diamond chain, equal to the sum of the
currents from $a_n$ to $b_n$ and to $c_n$, is thus found to be
\begin{align}
I(n\rightarrow n+1)=-(2J^2/\hbar\epsilon) \Im
[\langle\Psi_a(n)|{\bf
V}|\Psi_a(n+1)\rangle]. 
\end{align}
For a single propagating solution of the form (\ref{PSI}), Eqs.
(\ref{V}) and (\ref{spin}) yield
\begin{align}
I(n\rightarrow
n+1)&=-(4J^2/\hbar\epsilon)|A|^2
\bigl[\sin(q{\bar L})\bigl (\cos(\phi/2)c^2\nonumber\\
&+\mu n_z\sin(\phi/2)s^2 \bigr )\nonumber\\
&+\mu (n_x-n_y)\cos(q {\bar L})\cos(\phi/2)sc\bigr ].
\end{align}
It is easy to see that $I$ flips sign with $q$. When we have only
a pair of propagating modes, we thus concentrate on the one with
$I>0$.

\section{Results}

\begin{figure}[ht]
\begin{center}
\includegraphics[width=8 cm]{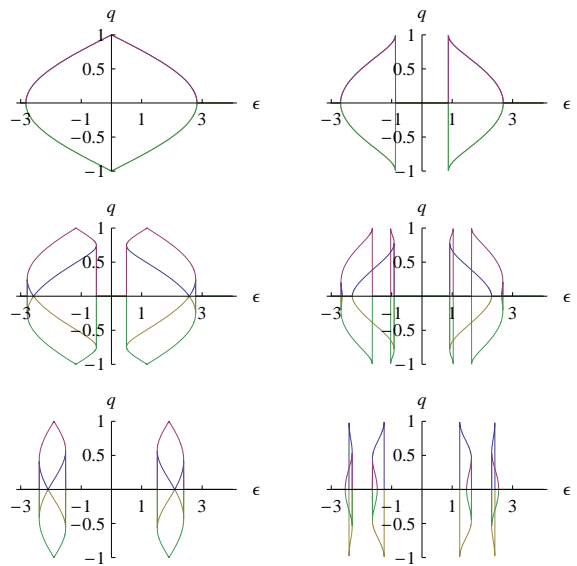}
\end{center}\vspace{-.3cm}
\caption{(Color online) The spectrum ($q$ versus $\epsilon$) of
the propagating solutions. Here, $J=1$ and the wave vector $q$ is
in units of $\pi{\bar L}$. Left: $\phi=0$. Right: $\phi=0.4\pi$.
Top to bottom: $\alpha=0,~0.2\pi,~0.4\pi$. The vertical lines
indicate boundaries at which the number of propagating solutions
changes. }\label{plot1}
\end{figure}

Figure \ref{plot1} shows the spectrum $\epsilon(q)$ of the propagating
solutions (real $q$'s), for several values of $\phi$ and $\alpha$. The left
column shows results for $\phi=0$, similar to Ref. \onlinecite{berc1}:
increasing $\alpha$ splits the energy band vertically, and changes
its width. Thus, the SOI can turn propagating waves into
evanescent ones, with complex $q$ (our figures show only the
solutions with real $q$). However, whenever the energy $\epsilon$
allows for real values of $q$, there exist four such values,
forming  pairs which move in opposite directions and have opposite
spins along $(1,-1,0)$. 

The situation becomes more interesting when we have both the SOI
and the AB flux. Adding only the latter (upper plot on the right hand side of
Fig. \ref{plot1}) creates
 a gap (i.e. evanescent states) around $\epsilon=0$. The degeneracy of
 the propagating solutions is not lifted, since the
two spin directions have exactly the same energies. As seen in the
right column in Fig. \ref{plot1}, increasing $\alpha$ at fixed
$\phi=0.4 \pi$ causes the splitting of each sub-band horizontally.

We next discuss the ballistic conductance of our device, $G$. For
an ideal conductor, this conductance is given by $G=(e^2/h)g$,
where $g$ is the number of right-moving (or left-moving)
propagating modes at a given energy. \cite{land,Imry,MV} This
formula clearly applies for the infinite periodic chain of
diamonds discussed here. Below we argue that the filtering effect
which we find also survives for a finite chain, under certain
conditions. As Fig. \ref{plot1} shows, at a given energy
$\epsilon$ one can encounter zero, two or four propagating
solutions.   The number $g$ can be read directly from Fig.
\ref{plot1}: on the left hand side of this figure, the number of
real $q$'s (both left-moving and right-moving) is always zero or
four, and thus $g=0$ or $2$. In contrast, the right hand side of
Fig. \ref{plot1} shows 0, 2 or 4 real $q$'s, i.e. $g=0,~1$ or $2$,
depending on the parameters $\epsilon,~\phi$ and $\alpha$.

We next consider electrons coming with arbitrary spin directions
from a reservoir at $-\infty$, with energy $\epsilon$ equal to
their chemical potential in that reservoir. For each electron, its
spinor will become a combination of the eigenmodes of the problem
inside the system. In fact, the same will happen to electrons
which enter into a {\it finite} but long chain from the left hand side:
their spinor within the chain will become a similar combination of the
four eigensolutions there, multiplied by some transmission coefficients.
When all four $q$'s have non-zero imaginary parts, all of these
modes are evanescent, and the wave function will decay to zero,
resulting with zero current. In that case there are no propagating
modes, and $g=0$. When all four $q$'s are real, i.e. $g=2$, the
incoming wave function is a combination of two right-moving modes,
and it has no definite spin. However, for $g=1$ the wave function
of the right moving electron is a linear combination of one
propagating and one evanescent modes. The latter will decay, and
the spinor will converge to that of the {\it single propagating
solution}, which has a {\it uniquely polarized spin}, see Eq.
(\ref{spin}). Without the AB flux, we always had $g=2$ or $g=0$.
For $\phi\ne 0$, we find regions of energy where $g=1$. Figure
\ref{cplt} shows contour plots of $g$ in the $\phi-\alpha$ plane,
for several values of $\epsilon$. As one can see, for energies
$\epsilon=-1.2J$ and $\epsilon=-2.4J$ there are large regions
where $g=1$. In these regions, the electron will have a well defined
polarization, which depends only on $\epsilon,~\phi$ and $\alpha$.

\begin{figure}[ht]
\begin{center}
\includegraphics[width=8cm]{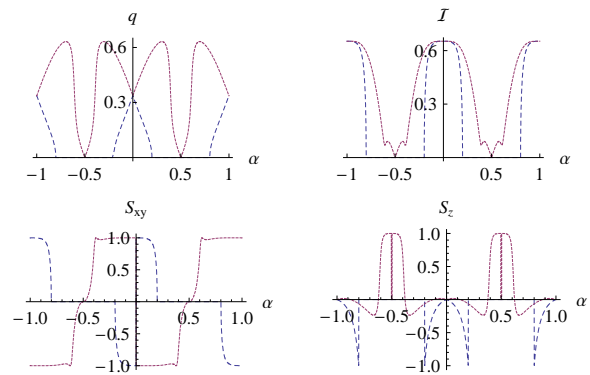}
\end{center}\vspace{-.3cm}
\caption{(Color online) Wave vectors $q$ (in units of $\pi{\bar
L}$), currents $I$, and spin components $S_{xy}$ and $S_z$, for
right-moving modes, as functions of the SO strength $\alpha$ (in
units of $\pi$), for $\epsilon/J=-2.4$ and $\phi=0.29\pi$. For
values of $\alpha$ at which $g=1$, the figures show only one mode
(dotted line). When $g=2$, the figures show two modes (dotted and
dashed lines).}\label{plso}
\end{figure}

\begin{figure}[ht]
\begin{center}
\includegraphics[width=8cm]{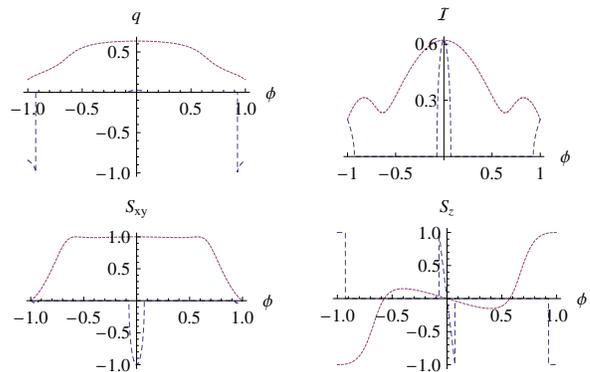}
\end{center}\vspace{-.3cm}
\caption{(Color online) Same as Fig. \ref{plso}, for
$\epsilon/J=-2.4$ and for fixed SO strength $\alpha=0.75\pi$, as
functions of $\phi$ (in units of $\pi$).}\label{plab}
\end{figure}

We next consider specific cuts through these contour plots. Figure
\ref{plso} shows  results as a function of $\alpha$ for fixed
energy $\epsilon/J=-2.4$ and AB phase $\phi=0.29\pi$. The plots
show only the ($g=1$ or $g=2$) right-moving modes ($I>0$). The
other propagating modes have opposite signs for $q$, $I$ and
$S_{xy}$. The dashed curves represent the second mode, which
arises only when $g=2$. For our purposes, we concentrate on the
regions where $g=1$, where one has only the dotted lines. The top
plots show the solutions for $q$ and the corresponding currents
$I$. The bottom plots show the spin components $S_{xy}$ and $S_z$.
The variation of $S_{xy}$ with $\alpha$ is striking: the spins of
the propagating electrons switch the sign of their in-plane
component with a small change of $\alpha$ near $\alpha=\pm0.5\pi$.
Note also the flipping of $S_{xy}$ as $\alpha$ crosses $\pm \pi$.
This flipping persists as $\phi$ increases, and the range with
$g=2$ near these points narrows. Figure \ref{plab} shows results
as a function of $\phi$, for the same energy, but at fixed SO
strength $\alpha=0.75\pi$. Clearly, even a relatively small AB
flux already yields a single right-moving propagating mode ($g=1$)
and therefore fully polarized spins. At small $\phi$, the
polarization starts close to the $(1,-1,0)$ direction, but it then
rotates towards the $z-$direction as $\phi$ increases towards $\pm
\pi$, and flips sign after crossing these points.

\begin{figure}[ht]
\begin{center}
\end{center}\vspace{-.3cm}
\caption{(Supplied separately) Contour plots of the ballistic
conductance (in units of $e^2/h$) in the $\phi-\alpha$ plane (the
AB phase $\phi$ and the SO strength $\alpha$ are in units of
$\pi$). The values $g=0,1,2$ are represented by dark, medium and
bright areas. The number above each plot is the energy $\epsilon$
(in units of $J$).}\label{cplt}
\end{figure}

\section{Discussion}

Given the above analysis, we may compare our system with that of
the single diamond, Ref. \onlinecite{hatano}. As we report
elsewhere, \cite{future} the single diamond generates fully
polarized electrons, along a controllable direction, whenever
$\sin^4\alpha=\cos^2(\phi/2)$ and for any $\epsilon$. Although
this condition is less restrictive than that given in Ref.
\onlinecite{hatano}, it is still much more restrictive than the
conditions we found above. The literature contains many other
proposals for spin filters, also based on the Rashba SOI. Usually,
these give only a partial polarization. Some of these devices also
require a large Zeeman field. In contrast, our filter can work at
a relatively low (and fixed) magnetic field (as apparently desired
technologically), so that the Zeeman energy is negligible. Note
also that both $\alpha$ and $\phi$ depend on the diamond size $L$,
and therefore one can choose a geometry which corresponds to the
available ranges of the magnetic field and the microscopic Rashba
parameters.

In real experiments it is not realistic to use an infinite chain
of diamonds. We now argue that under appropriate conditions it is
sufficient to use a finite chain, as long as it is longer than the
decay lengths of the evanescent modes. For the electrons coming in
from the left we don't need to worry about the details of the
connection between the incoming lead and the chain: even if some
of the electrons are reflected back into that lead, those which
are transmitted into the chain will split into a sum of the four
modes there, and when $g=1$ we still remain with fully polarized
electrons (although their overall amplitude may involve a
transmission factor with magnitude smaller than 1). The situation
on the right hand end of the chain is more delicate. Here we
should avoid reflections, since they may modify the outgoing
spinors and change their polarization.  A standard way to avoid
reflections is to use {\it adiabatic} contacts. This is usually
done for retaining the ballistic conductance of mesoscopic
devices. \cite{Imry} One way to avoid reflections is to have a
large leakage to the ground near the exit channel, so that only a
small fraction of electrons enter into the exit lead.

For our filter to be useful, one also needs to measure the outgoing
spins, or to relate the outgoing spin polarization to some
measurement of a voltage or a current. This issue is common to
many proposed filters, and it requires separate research. For the
present purposes, we mention just a few possibilities.
 First, one can follow the original proposal of Datta and
Das, \cite{2} and connect the right hand end of the device
adiabatically to a ferromagnetic lead, whose magnetization can be
tuned. The outgoing current will decrease with the angle between
the electron polarization and this magnetization.  Second, to
avoid connections to ferromagnets, one can also connect our filter
adiabatically to another such filter, with different parameters
which may block the polarized electrons coming from the first
filter.

Another way to test the spin polarization, is to couple one of the
$a$-nodes (Fig. \ref{1}) to a side quantum dot, that is in a Pauli
spin blockade region. \cite{ono} After a while, the side dot will
capture one of the polarized electrons, and this will block the
current (which contains electrons with the same polarization).
Changing the parameters will then change the spin direction of the
propagating electrons, and allow some current until the next
blocking occurs.

In conclusion, we propose a simple spin filter, which yields a
full polarization over a broad range of parameters. For given energy $\epsilon$ and magnetic flux $\phi$
(which need not be very large), the polarization of the outgoing electrons can be tuned
by varying the electric field which determines the SOI strength $\alpha$.

We acknowledge discussions with Joe Imry. AA and OEW
acknowledge the hospitality of NTT and of the ISSP, where this
project started, and support from the ISF and from the DIP.

\end{document}